\documentclass[preprint,prd,aps,nofootinbib]{revtex4} 
\usepackage{amssymb}
\usepackage{axodraw}
\usepackage{graphicx}

\begin{document}
 
\title {\large Lepton flavor violation in lopsided models\\
and a neutrino mass model}
 
\author{ Xiao-Jun Bi }
\affiliation{ Department of Physics, Tsinghua University, Beijing 100084,
People's Republic of China}
\email[Email: ]{bixj@mail.tsinghua.edu.cn}
 
\date{\today}

\begin{abstract}          

A widely adopted theoretical scheme to account for the neutrino oscillation
phenomena is the see-saw mechanism together with the ``lopsided''
mass matrices, which is generally realized in the framework of supersymmetric
grand unification. We will show that this scheme leads
to large lepton flavor violation at low energy if supersymmetry
is broken at the GUT or Plank scale. Especially, the branching ratio
of $\mu\to e\gamma$ already exceeds the present experimental limit.
We then propose a phenomenological model, which can account for
the LMA solution to the solar neutrino problem and at the same
time predict  branching ratio of $\mu\to e\gamma$  below the
present limit.

\end{abstract}
 
%\pacs{14.60.Pq}
\preprint{TUHEP-TH-02140}
 
\maketitle
 
\section { see-saw mechanism and ``lopsided'' structure }

The neutrino experiments show that the neutrino parameters have
two exotic while interesting features, \textit{i.e.},
the extreme smallness of the neutrino masses and the large size of
the neutrino mixing angles\cite{sk,k2k,sno}.
According to the recent analyses the
atmospheric neutrino oscillation favors the $\nu_\mu-\nu_\tau$
process with the best fit values\cite{atm}
\begin{equation}
\Delta m^2_{atm}= 2.5\times 10^{-3} eV^2,\ \
\sin^22\theta_{atm}=1.
\end{equation}
Among the four oscillation solutions for the solar
neutrino problem, the large mixing angle MSW (LMA) solution is
most favored, followed by the LOW and VAC solutions\cite{sma,lma,bahcall}.
The best fit values for the LMA solution are\cite{bahcall}
\begin{equation}
\Delta m^2_{sol}=5\times 10^{-5} eV^2,\ \ \tan^2\theta_{sol}=0.42.
\end{equation}
The same analysis excludes the small mixing angle (SMA) solution
at 3.7 $\sigma$ level.
%These two features show that the properties of neutrinos are 
%quite different from those of quarks and charged leptons.  
%These very differences may provide important hints useful  
%for solving the general flavor problem.

On the theoretical side, hundreds of neutrino mass
models have been constructed in 
the literature\cite{barr},
each trying to explain to a greater or lesser degree the two
afore-mentioned features.  A consensus has now
emerged that the see-saw mechanism seems to be the most natural
and economical way to account for the tiny neutrino masses. 

In the see-saw mechanism, the Standard Model (SM) is extended by
including the right-handed Majorana neutrinos, $\nu_R$.
Since $\nu_R$ are the SM gauge group, 
$SU(2)_W\times U(1)_Y$, singlets, their masses are not
protected by the SM gauge symmetry. The $\nu_R$ may
get masses at very high energy scale and may be much heavier
than the SM particles. Having both left- and right-handed
neutrinos and the $\nu_R$ being singlets, the neutrinos can 
have both Dirac mass terms, 
\begin{equation}
{\cal L}_D=-M_D \bar{\nu}_L\nu_R+h.c.\ ,
\end{equation} 
and Majorana mass terms, 
\begin{equation}
{\cal L}_M=-\frac{1}{2}M_R\nu_R^TC\nu_R+h.c.\ ,
\end{equation} 
with $C$ being the charge conjugate matrix.
Integrating out the heavy right-handed neutrinos, we get the
Majorana mass terms for the left-handed neutrinos, 
\begin{equation} 
{\cal L}_\nu=-\frac{1}{2}M_\nu\nu_L^TC\nu_L+h.c.\ , 
\end{equation}
with
\begin{equation}
\label{ss}
M_\nu=-M_DM_R^{-1}M_D^T\ \ .
\end{equation}
Since $M_R\gg M_D\sim M_{EW}$, we have $M_\nu$ is much smaller than
the electro-weak scale $M_{EW}$.

The see-saw mechanism is typically realized within the framework of a
supersymmetric (SUSY) grand unified theory (GUT), which adds further desirable
features including unification of the SM gauge
couplings at the GUT scale and avoidance of the SM hierarchy
problem. In an SO(10) GUT, see-saw mechanism is a natural outcome
of the group theory.

However, no generally accepted mechanism has yet been
put forth to explain the large neutrino mixing angles
until now\cite{barr}. The difficulty relies on the two facts
that (i) the neutrino spectrum exhibits large hierarchy, which
usually means small mixing among the neutrinos and (ii) in grand unified 
models the lepton and the quark mass matrices are closely related, 
which generally makes it difficult to accommodate
small quark mixing and large lepton mixing in one scheme.

An elegant idea proposed to explain the large neutrino mixing angle
is the so called ``lopsided'' structure\cite{albright,lopsid}. 
In this scheme the neutrino mass matrix, $M_\nu$, produces small mixing
according to the fact (i). However, the charged lepton
mass matrix, $M_L$, produces large mixing and the difficulty
relying on the fact (ii) is cleverly solved.
As we know, the
neutrino mixing is actually the mismatch between 
$M_L$ and $M_\nu$.
Diagonalizing $M_L$ and $M_\nu$ by
\begin{equation}
U_L^\dagger M_L U_R=\text{diag}(m_e,m_\mu,m_\tau)\ ,
\end{equation}
and
\begin{equation}
U_\nu^\dagger M_\nu U_\nu=\text{diag}(m_{\nu_1},m_{\nu_2},m_{\nu_3})\ ,
\end{equation} 
we have the neutrino mixing matrix
\begin{equation}  
\label{vmns}
V_{MNS}=U_L^\dagger U_\nu\ .
\end{equation}
So the large mixing in $U_L$ leads to large mixing
in the physical mixing matrix, $V_{MNS}$.

The ``lopsided'' structure works as follow.
In an SU(5) grand unified model, 
the left-handed charged leptons are in the
same multiplets as the CP conjugates of the right-handed down-type quarks,
and therefore $M_L$  is closely
related to the {\em transpose} of the mass matrix of the
down-type quarks, $M_{down}$. The two mass matrices
have the following approximate forms:
\begin{equation}
\label{lop}
M_L\ \sim\ \left( \begin{array}{ccc}
0 & 0 & 0 \\ 0 & 0 & \sigma \\ 0 & \epsilon & 1 \end{array} \right) m_D
\;\;\; \hbox{and} \;\;\;
M_{down}\ \sim \ \left( \begin{array}{ccc}
0 & 0 & 0 \\ 0 & 0 & \epsilon \\ 0 & \sigma & 1 \end{array} \right) m_D
\;\, ,
\end{equation}
respectively, with $\sigma \sim 1$, $\epsilon \ll 1$, and the
zeros representing  entries much smaller than $\epsilon$. For 
$M_L$, $\sigma$ controls the mixing between the second
and the third families of the left-handed leptons\footnote{Here we use the
convention that a left-handed doublet multiplies the Yukawa coupling matrix
from the left side while a right-handed singlet multiplies the matrix from
the right side.}, which greatly enhances $\theta_{atm}$,
while $\epsilon$ controls the
mixing between the second and the third families of the right-handed
leptons, which is not observable at low energy.  For the quarks
the roles of $\sigma$ and $\epsilon$ are reversed:  the small
$\mathcal{O}(\epsilon)$ mixing is in the left-handed sector, accounting for
the smallness of $V_{cb}$, while the large $\mathcal{O}(\sigma)$
mixing is in the right-handed sector, which is not observable.

A larger gauge group with SU(5) being its subgroup also has
the above property. 
Many realistic supersymmetric grand unified models have been built based 
on the ideas of see-saw mechanism and ``lopsided'' 
structure in the literature to account for the neutrino 
properties\cite{albright,lopsid}.
All such models have a definite 
prediction --- the lepton flavor violation (LFV) at low energy, 
which can be used to test this kind of models. We investigate
the LFV prediction in this kind of models.

\section{ Lepton flavor violation in supersymmetry}

In a supersymmetric model, the soft SUSY breaking terms may induce
large lepton flavor violation.
The possible LFV sources are the 
off-diagonal terms of the slepton mass matrices $(m^2_{\tilde{L}})_{ij}$,
$(m^2_{\tilde{R}})_{ij}$ and the trilinear couplings $A^L_{ij}$.
Present experimental bounds on the LFV processes give strong
constraints on such off-diagonal terms, with the strongest constraint
coming from Br($\mu\to e\gamma$) ($<1.2\times 10^{-11}$\cite{mueg}).
We have to find a mechanism to align the lepton and the scalar lepton
bases.  This is the so called SUSY flavor problem.

A generally adopted way to avoid these dangerous off-diagonal
terms is to impose universality constraints on the soft terms at
the SUSY-breaking scale, such as in the
gravity-mediated\cite{sgra} or gauge-mediated\cite{gmsb}
SUSY-breaking scenarios.  
Yet, even with the universality condition, off-diagonal
terms can be induced at lower energy scales through quantum
effects.  Such LFV effects
induced in the SUSY see-saw mechanism are given in the next
section. We first give the general analytic expressions for the
branching ratios of the LFV processes, $l_i\to l_j\gamma$.

\begin{center}
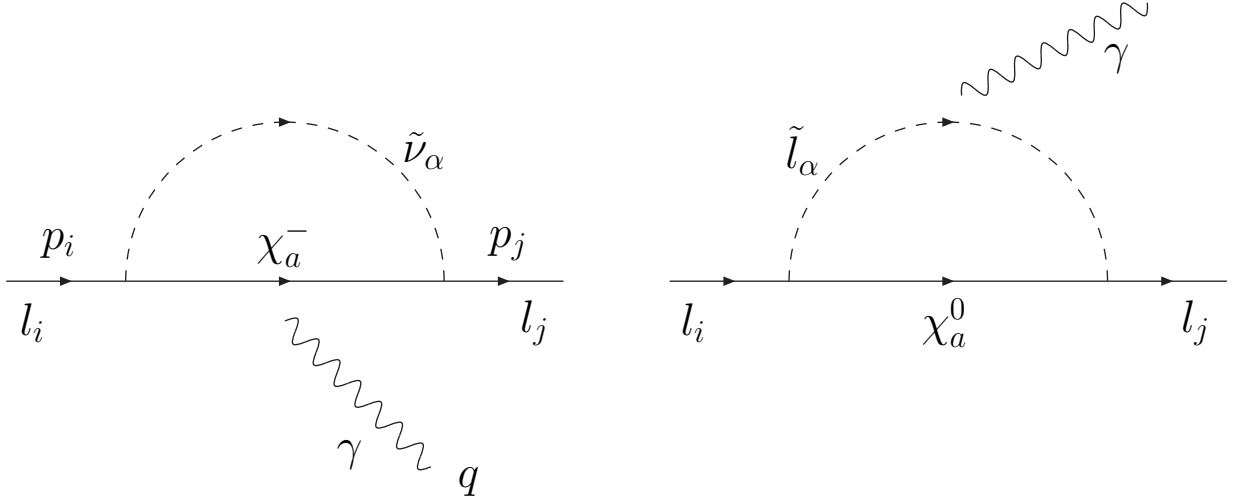
\begin{figure}
\begin{picture}(500,250)(20,50)
\ArrowLine(20,200)(65,200)
\ArrowLine(65,200)(185,200)
\ArrowLine(185,200)(230,200)
\ArrowLine(270,200)(315,200)
\ArrowLine(315,200)(435,200)
\ArrowLine(435,200)(480,200)
\DashArrowArcn(125,200)(60,180,0){4}
\DashArrowArcn(375,200)(60,180,0){4}
\Photon(125,185)(180,130){4}{7}
\Text(150,135)[]{\Large $\gamma$}
\Text(170,250)[l]{\Large $\tilde{\nu}_\alpha$}
\Text(30,185)[]{\Large $l_i$}
\Text(220,185)[]{\Large $l_j$}
\Text(125,215)[]{\Large $\chi^-_a$}
\Text(40,215)[]{\Large $p_i$}
\Text(210,215)[]{\Large $p_j$}
\Text(195,125)[]{\Large $q$}

\Photon(380,270)(450,305){4}{7}
\Text(440,285)[]{\Large $\gamma$}
\Text(328,250)[r]{\Large $\tilde{l}_\alpha$}
\Text(280,185)[]{\Large $l_i$}
\Text(470,185)[]{\Large $l_j$}
\Text(375,185)[]{\Large $\chi^0_a$}
\end{picture}

\vspace*{-1.5cm}
\caption{\label{fig1} Feynman diagrams for the process $l_i \to l_j\gamma$
via the exchange of a chargino (left) and via a neutralino (right).}
\end{figure}
\end{center}

The LFV decay,  $l_i\to l_j\gamma$, occurs through the 
photon-penguin diagrams
shown in FIG. \ref{fig1}. The amplitude for the 
processes takes the general form
\begin{equation}
\label{amp}
M=em_i \bar{u}_j(p_j)i\sigma_{\mu\nu}q^\nu
(A^{ij}_LP_L+A^{ij}_RP_R)u_i(p_i)\epsilon^\mu (q)\ \ .
\end{equation}
The contribution from neutralino exchange gives
\begin{eqnarray}
A_L^{(n)}&=&-\frac{1}{32\pi^2}(\frac{e}{\sqrt{2}\cos\theta_W})^2\frac{1}
{m_{\tilde{l}_\alpha}^2}\left[{B^{j\alpha a}}^*B^{i\alpha
a}F_1(k_{\alpha a}) +\frac{m_{\chi_a^0}}{m_i}{B^{j\alpha a}}^*
A^{i\alpha a}F_2(k_{\alpha a})\right]\ ,\\
A_R^{(n)}&=&A_L^{(n)}\ (B\leftrightarrow A)\ ,
\end{eqnarray}
where
\begin{eqnarray}
F_1(k)&=&\frac{1-6k+3k^2+2k^3-6k^2\log k}{6(1-k)^4}\,\\
F_2(k)&=&\frac{1-k^2+2k\log k}{(1-k)^3}\ \ ,
\end{eqnarray}
with $k_{\alpha a}={m_{\chi_a^0}^2}/{m_{\tilde{l}_\alpha}^2}$. $A$ and $B$ are
the lepton--slepton--neutralino coupling vertices given by
\begin{eqnarray}
A^{i\alpha
a}&=&\left(Z_{\tilde{L}}^{i\alpha}(Z_N^{1a}+Z_N^{2a}\cot\theta_W)
-\cot\theta_W\frac{m_i}{M_W\cos\beta}Z_{\tilde{L}}^{(i+3)\alpha}
{Z_N^{3a}}\right) \ ,\\
B^{i\alpha
a}&=&-\left(2Z_{\tilde{L}}^{(i+3)\alpha}{Z_N^{1a}}^*+\cot\theta_W
\frac{m_i}{M_W\cos\beta}Z_{\tilde{L}}^{i\alpha}{Z_N^{3a}}^*\right)
\; ,
\end{eqnarray}
where $Z_{\tilde{L}}$ is the $6\times 6$ slepton mixing matrix and
$Z_N$ is the neutralino mixing matrix.
The corresponding contribution coming from chargino exchange is
\begin{eqnarray}
A_L^{(c)}&=&\frac{g_2^2}{32\pi^2}{Z_{\tilde{\nu}}^{i\alpha}}^*
Z_{\tilde{\nu}}^{j\alpha}
\frac{1}{m_{\tilde{\nu}_\alpha}^2}\left[ Z^-_{2a}{Z^-_{2a}}^*
\frac{m_im_j}{2M_W^2\cos^2\beta}
F_3(k_{\alpha a})\right. \nonumber \\
&&\left.
+\frac{m_{\chi^-_a}}{\sqrt{2}M_W\cos\beta}Z^+_{1a}Z^-_{2a}
\frac{m_j}{m_i} F_4(k_{\alpha a})\right] \ ,\\
\label{imp}
A_R^{(c)}&=&\frac{g_2^2}{32\pi^2}{Z_{\tilde{\nu}}^{i\alpha}}^*
Z_{\tilde{\nu}}^{j\alpha} \frac{1}{m_{\tilde{\nu}_\alpha}^2}\left[
Z^+_{1a}{Z^+_{1a}}^*F_3(k_{\alpha a})
+\frac{m_{\chi^-_a}}{\sqrt{2}M_W\cos\beta}{Z^+_{1a}}^*{Z^-_{2a}}^*
F_4(k_{\alpha a})\right] \; ,
\end{eqnarray}
where
\begin{eqnarray}
F_3(k)&=&\frac{2+3k-6k^2+k^3+6k\log k}{6(1-k)^4} \ ,\\
F_4(k)&=&\frac{3-4k+k^2+2\log k}{(1-k)^3}\ \ ,
\end{eqnarray}
with $k_{\alpha a}={m_{\chi_a^-}^2}/{m_{\tilde{\nu}_\alpha}^2}$.
$Z_{\tilde{\nu}}$ is the sneutrino mixing matrix, while
$Z^+$ and $Z^-$ are the chargino mixing matrices.

The branching ratio for $l_i\to l_j\gamma$ is given by
\begin{equation}
\text{Br}(l_i\to l_j\gamma)=\frac{\alpha_{em}}{4}m_i^5(|A^{ij}_L|^2+|A^{ij}_R|^2)/
\Gamma_i \ \ ,
\end{equation}
where $\Gamma_i$ is the width of $l_i$.
To identify the parameter dependence one may use the mass
insertion approximation\cite{casas}, which yields,
for large $\tan\beta$,
\begin{equation}
\label{appbr}
\text{Br}(l_i\to l_j\gamma)\ \sim\ \frac{\alpha^3}{G_F^2}
\frac{ [(m^2_{\tilde{L}})_{ij}]^2 }{m_s^8}\tan^2\beta\ \ ,
\end{equation}
where $m_s$ represents the common slepton mass. 
We can see that the supersymmetric
contribution to Br($l_i\to l_j\gamma$) is proportional to
$\tan^2\beta$ and to the amount of the off-diagonal terms in 
the slepton mass matrix.

\section{ radiatively produced LFV in see-saw mechanism}

Although the soft terms are universal at the GUT (or Plank) scale,
off-diagonal soft terms may be radiatively
produced in the see-saw mechanism. Especially, 
if the charged lepton mass matrix
is ``lopsided'', the radiatively produced LFV effects are large
enough to be observed. We will show this below.

At the energy scales between $M_R$ and $M_{GUT}$,
the superpotential of the lepton sector is given by
\begin{equation}
W=Y_N^{ij} \hat{H}_2\hat{L}_i\hat{N}_j+Y_L^{ij}\hat{H}_1\hat{L}_i\hat{E}_j
+\frac{1}{2}M_R^{ij}\hat{N}_i\hat{N}_j+\mu \hat{H}_1\hat{H}_2\ \ ,
\end{equation}
where $Y_N$ and $Y_L$ are the neutrino and
charged lepton Yukawa coupling matrices, respectively.
In general, $Y_N$ and $Y_L$ can not be diagonalized simultaneously.
This bases misalignment can lead to lepton flavor violation, similar 
to the quark sector.
This LFV effects can transfer to the soft terms through quantum effects and
induce non-diagonal terms below the GUT scale.
This is clearly shown by the following
renormalization group equation (RGE) for $m_{\tilde{L}}^2$, 
which gives the dominant contribution to low
energy LFV processes,
\begin{eqnarray}
\mu\frac{dm_{\tilde{L}}^2}{d\mu}&=&\frac{2}{16\pi^2}
\left[-\Sigma c_ig_i^2M_i^2+
\frac{1}{2}[Y_NY_N^\dagger m_{\tilde{L}}^2+m_{\tilde{L}}^2Y_NY_N^\dagger]
+\frac{1}{2}[Y_LY_L^\dagger m_{\tilde{L}}^2+m_{\tilde{L}}^2Y_LY_L^\dagger]
\right.\nonumber\\
&&\left.+Y_Lm_{\tilde{E}}^2Y_L^\dagger+m_{H_D}^2Y_LY_L^\dagger+E_AE_A^\dagger
+Y_N m_{\tilde{N}}^2Y_N^\dagger+m_{H_U}^2Y_NY_N^\dagger+N_AN_A^\dagger\right]
\;\;\; .
\end{eqnarray}
Here $E_A=A^L\cdot Y_L$ and $N_A=A^N\cdot Y_N$, while $g_i$ and $M_i$
are the gauge coupling constants and gaugino masses, respectively.

$Y_L$ and $Y_N$ can be diagonalized by bi-unitary rotations
\begin{equation}
Y^\delta_L=U_L^\dagger Y_L U_R\  ,\ \
Y^\delta_N=V_L^\dagger Y_N V_R\  ,
\end{equation}
respectively.  Lepton flavor mixing is determined by the
matrix $V_D$, the analog to $V_{KM}$ in the quark
sector, defined by
\begin{equation}
\label{vd} V_D=U_L^\dagger V_L\ .
\end{equation}
%We see that $V_D$ is determined by the left-handed mixing of the Yukawa
%coupling matrices $Y_L$ and $Y_N$. It 
We see that $V_D$ only exists above
the energy scale $M_R$. It is different from the MNS matrix
$V_{MNS}$ in Eq. (\ref{vmns}).

Then running the RGEs between 
$M_{GUT}$ (where the initial soft terms are universal)
and $M_R$ (where $\nu_R$ decouples and no LFV interactions below)
leads to the flavor mixing off-diagonal terms. 
On the basis where $Y_L$ is diagonal, 
the off-diagonal terms of $m_{\tilde{L}}^2$ can be approximately given as, 
\begin{eqnarray}
\left(\delta m_{\tilde{L}}^2\right)_{ij}
&\approx &\frac{1}{8\pi^2}
(Y_NY_N^\dagger)_{ij} (3+a^2)m_0^2
\log\frac{M_{GUT}}{M_R}\nonumber\\
\label{dm}
&\approx&\frac{1}{8\pi^2}
(V_D)_{i3}(V_D^*)_{j3}
Y_{N_3}^2(3+a^2)m_0^2\log\frac{M_{GUT}}{M_R} \ \ ,
\end{eqnarray}
where, assuming the three generations' Yukawa couplings in $Y_N$ are hierarchical,
only the third generation's Yukawa coupling, $Y_{N_3}$, is retained.
The `$a$' is the universal trilinear coupling given by $A_0=a m_0$, 
and $m_0$ is the universal slepton mass at $M_{GUT}$.

Eq. (\ref{dm}) clearly shows that the mixing matrix $V_D$
determines $\delta m_{\tilde{L}}^2$.
The ``lopsided'' models predicts big mixing in $U_L$, 
and therefore big mixing in $V_D$, which 
finally leads to observable LFV effects.

\section{Numerical results}

The precise results are obtained by solving the coupled
RGEs numerically. The RGEs below $M_R$ are the set of equations
for MSSM, while above $M_R$ the equations must be extended by
including $\nu_R$ and corresponding scalar partners. The
details for solving the equations are given in Ref. \cite{bi1}.

For the process $\tau\to \mu\gamma$, its branching ratio is 
approximately proportional to $|{(V_D)}_{23} {(V_D)}_{33}|^2$. 
This quantity is quite model
independent since all the ``lopsided'' models give a large, 
near maximal, 2-3 mixing.
%Assuming $\theta_{23}$ a large region $\sim 20^\circ - 70^\circ $ we have
%$|{(V_D)}_{23} {(V_D)}_{33}|\sim (0.5\sim 0.3)$.
%with $V_{23} \sim V_{33} \sim \frac{1}{\sqrt{2}}$ in the lopsided models. 
Thus we can give a quite definite prediction for this process.

\begin{figure}
\includegraphics[scale=0.6]{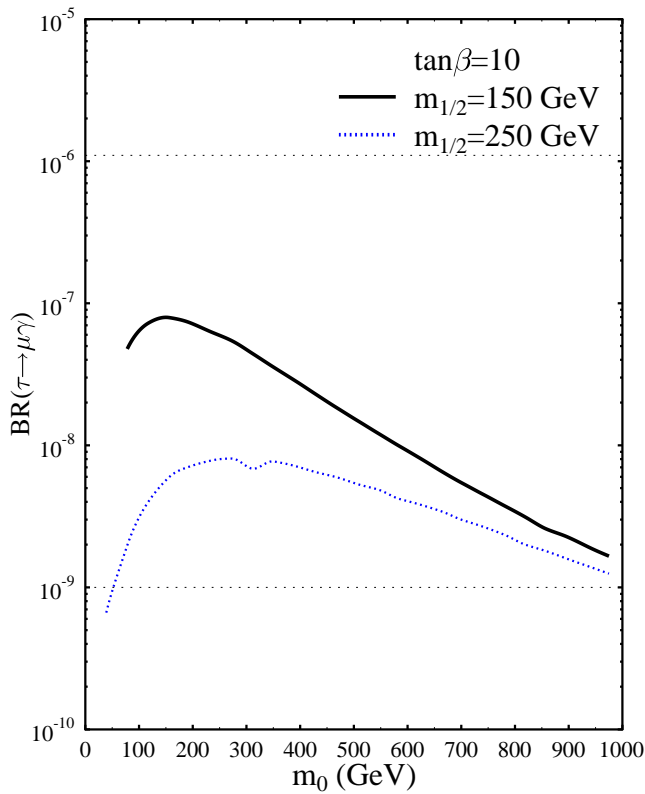}
\caption{\label{fig2}
Branching ratio of $\tau\rightarrow\mu\gamma$ as a function of $m_0$ for
$A_0=m_0$, $\tan\beta=10$ and $m_\frac{1}{2}=150GeV,\ 250GeV$.
The dotted lines are the present upper bound and the expected
sensitivity.}
\end{figure}
 
Br($\tau\to\mu\gamma$) is plotted in FIG. \ref{fig2} for a typical set
of SUSY parameters.
We notice that in a quite large parameter space
the process $\tau\to \mu\gamma$, induced in  supersymmetric see-saw
mechanism, is below the present experimental bound, 
$1.1\times 10^{-6}$\cite{pdg}, while, will be 
detected in the future experiment if the expected sensitivity
can reach down to $10^{-9}$\cite{ellis}.
In our calculation the SUSY parameters are constrained by
the $g_\mu-2$ anomaly\cite{g-2}, so $m_\frac{1}{2}$ can not
be too large.

The branching ratio of $\mu\to e\gamma$ is approximately proportional to
$|{(V_D)}_{13} {(V_D)}_{23}|^2$. 
The element ${(V_D)}_{13}$ seems
quite model-dependent. However, under the following observations
and assumptions, we find that a general 
prediction of ${(V_D)}_{13}$ in this kind of models is possible\cite{bi2}.

First, we assume that $Y_N$ has a similar hierarchical
structure as the Yukawa coupling matrix of the up-type quark, $Y_u$.
In SO(10) grand unified models, the simplest symmetry
breaking mechanism leads to $Y_N=Y_u$. Since the see-saw mechanism
is usually realized in an SO(10) grand unified model,
this assumption is quite general.
$Y_u$ is constrained by the values of up-type quark masses
and the CKM mixing angles. By our second assumption
that there is no accidental cancellation between the 
mixing matrices for the up- and down-type quarks leading to
small CKM mixing, we then have
\begin{equation}
\theta_u^{13}\lesssim V_{td}\sim 0.008\ ,
\end{equation}
with $\theta_u^{13}$ and $ V_{td}$ being the 1-3 mixing
angle produced by $Y_u$ and the 3-1 element of the CKM matrix.
We thus expect that the 1-3 mixing angle produced by $Y_N$,
 ${\theta}_N^{13}$, is of the same order of magnitude as $\theta_u^{13}$.
Then we have ${\theta}_N^{13}\lesssim 0.008$. Analogously, we
have $\sin{\theta}_N^{23}\lesssim V_{ts}\cong 0.04$. 
Third, we observed that in most 
models $m_\tau$ and $m_\mu$ got their masses mainly from the 2-3
block of the lepton mass matrix. The elements in the first row
and the first column of $M_L$ are constrained by $m_e$.
By this structure, as given in Eq. (\ref{lop}), one finds 
that\cite{barr,frit}
\begin{equation}
\sin{\theta}_{12}\sim \sqrt{m_e/m_\mu}\cong 0.07\ ,
\end{equation}   
and
\begin{equation} 
\sin{\theta}_{13}\approx m_\mu/m_\tau 
\sin{\theta}_{12}\ll\sin{\theta}_{12}\ ,
\end{equation}
with ${\theta}$ being mixing angles in $U_L$.
Finally, taking into account that ${\theta}_{23}\sim \mathcal
O(1)$ in ``lopsided'' models  we get
\begin{eqnarray}
\label{vd13}
({V_D})_{13} & \approx & 
\sin{\theta}_{12} \sin{\theta}_{23} \approx 0.05\ \ ,\\
\label{vd23}
({V_D})_{23} & \approx &  
-\sin{\theta}_{23} \approx -0.71\ \ .
\end{eqnarray}
Thus the angles in $U_L$ alone can determine $({V_D})_{13}$
and $({V_D})_{23}$.  This conclusion certainly depends on the assumed
forms of $Y_L$ and $Y_N$; nonetheless, it is correct
in most published ``lopsided'' models\cite{albright,lopsid}, 
which can be explicitly checked. 
In fact our assumptions are implied in these models.

Actually the above assumptions can be relaxed. 
Since Eq. (\ref{vd13}) is one term, the dominant one
here, of the full expressions for $({V_D})_{13}$, unless there is strong
cancellation among these terms, do we always have 
$({V_D})_{13}$ to be $\mathcal{O}(0.05)$ or larger.

\begin{figure}
\includegraphics[scale=0.6]{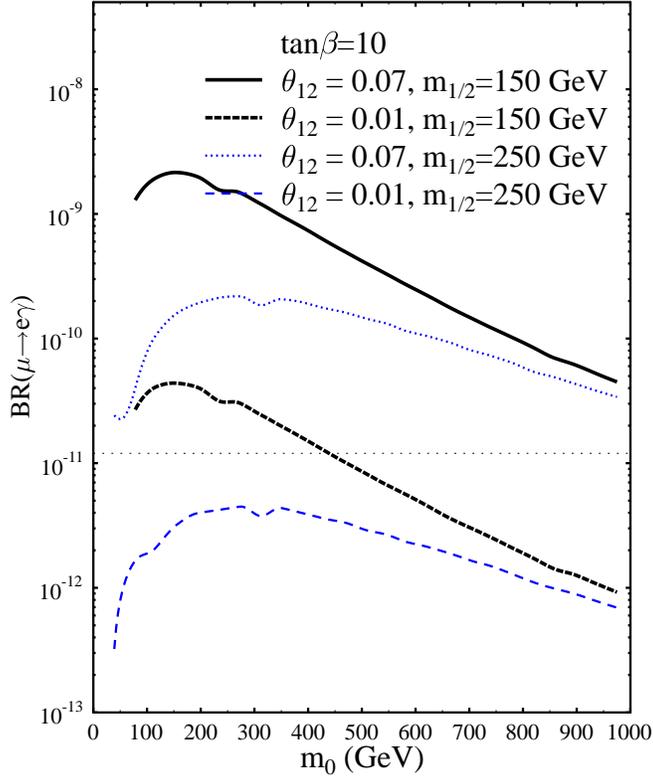}
\caption{\label{fig3}
Br($\mu\to e\gamma$) as a function of $m_0$ for $\tan\beta=10$ and
$m_{1/2}=150\, GeV, 250\, GeV$.
$A_0=m_0$ and $\mu > 0$ are assumed.
$\theta_{12}$ is the mixing angle between the $1^{st}$ \&
$2^{nd}$ generations in $U_L$. The horizontal dotted line
is the present experimental limit,
$1.2\times 10^{-11}$\cite{mueg}. }
\end{figure}

In FIG. \ref{fig3} we give our numerical result for Br($\mu\to e\gamma$). 
Taking  $\tan\beta=10$ 
and $\theta_{12}=0.07$ as the typical value
of the mixing angle between the $1^{st}$ and the $2^{nd}$
generations in $U_L$, we find that the predicted Br($\mu\to e\gamma$)
has already exceeded the present upper bound,
$1.2\times 10^{-11}$\cite{mueg}. The other set of curves
are for $\theta_{12}=0.01$ (corresponding to
${(V_D)}_{13}=0.007$).
In this case Br($\mu\to e\gamma$) may be below the experimental
limit. 

So large Br($\mu\to e\gamma$)
is because of the large mixing angle $\theta_{23}$, which features 
the ``lopsided'' model and gives satisfied solution to large neutrino
mixing. However, large $\theta_{23}$ enhances
both ${(V_D)}_{13}$ and ${(V_D)}_{23}$, as given in Eqs. (\ref{vd13}) 
and (\ref{vd23}), leading to too large Br($\mu\to e\gamma$). 
This is really a dilemma. Another shortage
of the ``lopsided'' model is that it generally predicts SMA or VAC solution
to the solar neutrino problem, which are disfavored by present
data. A recent work in Ref. \cite{albright} predicts the LMA solution
by ``lopsided'' structure.
However, fine tuning to some extend is needed in this model.
In next section we propose a new structure for $M_L$,
which can solve the above problems simultaneously.
The structure predicts very small ${(V_D)}_{13}$ while,   
at the same time, LMA solution to the solar neutrinos.

\section{ a new neutrino mass model}

Assuming $Y_N$ is nearly diagonal we give
\begin{equation}
\label{form}
M_L=\left( \begin{array}{ccc} 0 & \delta & \sigma \\ -\delta & 0 & 1-\epsilon\\
0 & \epsilon & 1 \end{array}\right) m,\ \ 
\text{with}\ \
\sigma\sim \mathcal{O}(1),\ \delta\ll\epsilon\ll 1\ .
\end{equation}
Taking 
\begin{equation}
\label{para} \delta=0.00077,\ \epsilon=0.12,\ \text{and}\ \
\sigma=0.58
\end{equation}
we can  obtain the correct mass ratios
$m_e/m_\mu$, $m_\mu/m_\tau$ and 
predict the neutrino mixing parameters as
\begin{equation} \sin^22\theta_{atm}=0.998, \tan^2\theta_{sol}=0.42\
\text{and}\ U_{e3}=-0.0054 .
\end{equation}

The notable feature of form (\ref{form}) compared with the usual
``lopsided'' models is the $\mathcal{O} (1)$ element
$\sigma$. Both the 
$(2,3)$ and $(1,3)$ elements in $M_L$ are large, 
naturally leading to large
mixing angles, $\theta_{23}$ and $\theta_{12}$. 
%The set of parameters lead to maximal atmospheric neutrino
%mixing and the large solar neutrino mixing, corresponding to
%the best fit value in the LMA solution.
The prediction of 
$U_{e3}=-0.0054$ is non-trivial, since the three parameters
are fixed by the lepton mass ratios and one neutrino mixing
angle.  It thus provides a test of our model.

Diagonalizing $M_L$ analytically we can express
$U_{e3}$ as
\begin{equation}
\label{ue3}
U_{e3}\cong\frac{m_e}{m_\mu}\cdot U_{\mu 3}/\tan\theta_{sol}\ .
\end{equation} 
This prediction that $U_{e3}$ is proportional to $m_e/m_\mu$ is unique.
Usually  $U_{e3}$ is predicted to be proportional to $\sqrt{m_e/m_\mu}$.
Our model gives very small  $U_{e3}$ value. 
Another interesting example which also gives quite small $U_{e3}$
is in Ref. \cite{xing}, which predicts 
$U_{e3}=\sqrt{\frac{m_e}{m_\tau}}U_{\mu 3}$.
However, this model predicts $\theta_{sol}\approx \pi/4$, which is
excluded by present data.

\begin{figure}
\includegraphics[scale=0.6]{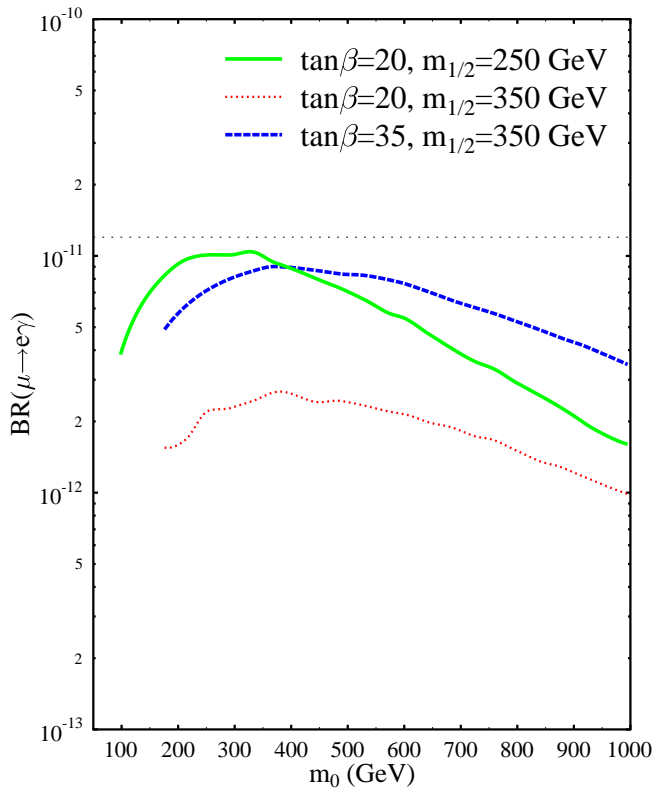}
\caption{\label{fig4}
$Br(\mu\to e\gamma)$ predicted by our model as a function of $m_0$ for
$\tan\beta=20$, $m_{1/2}=250\, GeV, 350\, GeV$
and $\tan\beta=35$, $m_{1/2}=350\, GeV$.
 The horizontal dotted line
is the present experimental limit,
$1.2\times 10^{-11}$\cite{mueg}. }
\end{figure}

The prediction of Br($\mu\to e\gamma$) by our model 
is plotted in FIG. \ref{fig4}. In most
parameter space our model predicts  Br($\mu\to e\gamma$)
below the present experimental limit, while large
enough to be detected in the next generation experiment\cite{nexp}.

\section { Summary and Conclusions }

A quite popular theoretical scheme to explain the atmospheric
and solar neutrino experiments is the see-saw mechanism
together with the ``lopsided'' charged lepton mass matrix.
This scheme is generally realized in the framework of 
supersymmetric grand unification. 
Our analysis shows that such a structure
may predict big lepton flavor violation at low energy.
The process $\tau\to \mu\gamma$ is quite promising to
test whether there is a large mixing in the charged lepton sector, as
predicted by ``lopsided'' models. In most SUSY parameter space this
process will be detected in the next generation experiment.
The ``lopsided'' models also make model-insensitive prediction about
the process $\mu\rightarrow e\gamma$. However,
the branching ratio of 
$\mu\rightarrow e\gamma$ predicted by these
models generally exceeds the present experimental limit. 
An extended ``lopsided'' form of the charged lepton mass 
matrix is then proposed to
solve this problem.
The new structure can produce maximal 2-3 mixing, 
large 1-2 mixing while very 
small 1-3 mixing in the 
lepton sector. Br($\mu\rightarrow e\gamma$) is thus suppressed 
below the present experimental limit. The LMA solution for the solar 
neutrino problem is naturally produced.

\begin{acknowledgments}

This work is supported by the National Natural Science Foundation
of China under the grand No. 10105004.

\end{acknowledgments}

\end{document}